\documentstyle[twocolumn,aps,epsfig,floats]{revtex}
\begin{document}

\twocolumn[\hsize\textwidth\columnwidth\hsize\csname %
@twocolumnfalse\endcsname

\title{Spin-lattice relaxation in the mixed state of
the high-$T_c$ cuprates:\\ electronic spin-flip scattering versus
spin-fluctuations}
\author{Dirk K. Morr}
\address{Theoretical Division, MS B262, Los Alamos National
Laboratory, Los Alamos, NM 87545 }
\date{\today}
\draft \maketitle
\begin{abstract}
Recent experimental and theoretical studies have established that
the spin-lattice relaxation rate, $1/T_1$, measured in nuclear
magnetic resonance (NMR) experiments is a site-sensitive probe for
the electronic spectrum in the mixed state of the high-$T_c$
cuprates. While some groups suggested that $1/T_1$ is solely
determined by spin-flip scattering of BCS-like electrons, other
groups stressed the importance of antiferromagnetic
spin-fluctuations. We show that these two relaxation mechanisms
give rise to a {\it qualitatively} different temperature and
frequency dependence of $^{17}$O $1/T_1$. A comparison of our
results with the experimental $^{17}$O $1/T_1$ data provides support for
a relaxation mechanism dominated by antiferromagnetic
spin-fluctuations.

\end{abstract}


]

Recently the temperature and frequency dependence of the $^{63}$Cu
and $^{17}$O spin-lattice relaxation rate, $1/T_1$, in the mixed
state of YBa$_2$Cu$_3$O$_{6+x}$ (YBCO) has been the focus of
intense experimental efforts  \cite{Cur98,Cur00,Mil99,Hal00}. It
was first shown by Curro {\it et al.}~(CMHS) \cite{Cur98,Cur00},
and subsequently by Halperin {\it et al.} \cite{Hal00} that
$^{17}$O $1/T_1$ in the mixed state of YBa$_2$Cu$_3$O$_7$
increases with increasing resonance frequency, i.e., with
decreasing distance from the vortex core. Preliminary results of
Milling {\it et al.}~\cite{Mil99} show a qualitatively similar
behavior for the $^{63}$Cu relaxation rate in YBCO. CMHS were the
first to point out that this frequency dependence of $1/T_1$
arises from the presence of a supercurrent, which gives rise to a
site-specific density of states, and consequently a site-specific
$1/T_1$. This interpretation receives strong support from the
experimental observation that the frequency dependence of $1/T_1$
disappears above the melting temperature of the vortex lattice.

Several theoretical models \cite{Tak99,Morr00a,Wor00}  have been suggested
to explain the experimentally observed
frequency dependence of $1/T_1$. While it was argued in
Refs.~\cite{Tak99,Wor00} that $1/T_1$ for $^{63}$Cu arises solely
from electronic spin-flip scattering of BCS-type electrons, Morr
and Wortis (MW) \cite{Morr00a} proposed that antiferromagnetic
spin-fluctuations provide the dominant contribution to the
$^{63}$Cu spin-lattice relaxation rate. MW also conjectured that
the combined temperature and frequency dependence of $1/T_1$ can
distinguish between relaxation due to electronic spin-flip (ESF)
scattering and relaxation due to antiferromagnetic
spin-fluctuations (ASF).

In this communication we show that due to the specific form of the
$^{17}$O hyperfine coupling, the temperature and frequency
dependence of the $^{17}$O spin-lattice relaxation rate for ESF
and ASF relaxation is {\it qualitatively} different. In
particular, we demonstrate that for the ESF mechanism, $1/T_1$
increases monotonically with temperature and possesses a large
distribution of values, for a given resonance frequency, close to
the vortex core. In contrast, for the ASF relaxation mechanism, we
obtain a well-defined relation between $1/T_1$ and resonance
frequency, and a non-monotonic dependence of $1/T_1$ on
temperature. Though a comparison of our theoretical results with
the available experimental data suggests that the spin-lattice
relaxation rate in the mixed state is dominated by the ASF
mechanism, further experiments are required to test the
predictions presented in this paper. To the extent that future
experiments confirm our conclusion, it immediately implies that
the relaxation rate of $^{63}$Cu is as well dominated by
antiferromagnetic spin fluctuations due to its even more favorable
hyperfine coupling.

We begin by shortly reviewing the theoretical models for an ESF
and ASF based spin-lattice relaxation rate; a detailed account is
given in Refs.~\cite{Tak99,Morr00a,Wor00}. For ESF relaxation \cite{com4}, we
consider the model of Ref.~\cite{Wor00}, whose results
are in good qualitative with those presented
in Ref.\cite{Tak99}.

The general expression for the $^{17}$O spin-lattice relaxation
rate, $1/T_1$, with an applied field parallel to the $c-$axis is
given by
\begin{equation}
{1 \over T_1 T} = { k_B \over 2 \hbar} (\hbar^2 \gamma_n
\gamma_e)^2 { 1 \over N} \sum_{\bf q} F_{\rm O}({\bf q})
\lim_{\omega \rightarrow 0} { \chi''({\bf q}, \omega) \over
\omega}  \ , \label{T1T}
\end{equation}
where
\begin{equation}
F_{{\rm O}(2)}(q) = C^2 \cos(q_x/2)^2 \quad F_{{\rm O}(3)}(q) =
C^2 \cos(q_y/2)^2\ , \label{form}
\end{equation}
$C$ is the $^{17}$O transferred hyperfine coupling constant, and
O(2)[O(3)] refers to the oxygen nucleus which is located between
two Cu nuclei along the x(y)-axis.

Within the ESF approach the spin-susceptibility in the
superconducting state, neglecting final state effects, is given by
$\chi=\Sigma$, where
\begin{eqnarray}
\Sigma({\bf q}, i \omega_n) &=& - T \sum_{{\bf k},m} \ \Big\{
G({\bf k}, i\Omega_m) G({\bf k+q}, i\Omega_m+i\omega_n) \nonumber
\\ & & + F({\bf k}, i\Omega_m) F({\bf k+q}, i\Omega_m+i\omega_n)
\Big\} \ ,
\label{Sigma}
\end{eqnarray}
and $G$ and $F$ are the normal and anomalous Green's functions
\begin{eqnarray}
G({\bf k}, i\Omega_m) &=& { v^2_{\bf k} \over i\Omega_m - E_{\bf
k} } + { u^2_{\bf k} \over i\Omega_m + E_{\bf - k} } \nonumber \ ;
\\ F({\bf k}, i\Omega_m) &=&  -u_{\bf k} v_{\bf k} \left\{ {1
\over i\Omega_m - E_{\bf k} } - { 1 \over i\Omega_m + E_{\bf - k}
} \right] \ .
\label{GF}
\end{eqnarray}
In the limit that the supercurrent momentum, ${\bf p}_s$, varies
on a length scale larger than $1/k_F$, the dispersion of the
fermionic quasi-particles, $E_{\bf k}$, up to linear order in
$p_s$ is in semiclassical approximation given by \cite{Tink80}
\begin{equation}
E_{\bf k} = \sqrt{  \epsilon_{\bf k}^2 + |\Delta_{\bf k}|^2} +
{\bf v}_F({\bf k})\cdot{\bf p}_s \ ,
\label{qp_disp}
\end{equation}
where $\epsilon_{\bf k}$ is the electronic normal state
dispersion, $\Delta_{\bf k}=\Delta_0 \ (\cos(k_x) - \cos(k_y))/2$
is the $d$-wave gap, and ${\bf v}_F({\bf k})=\partial
\epsilon_{\bf k} /\partial {\bf k}$. Following MW, we neglect the
electronic Zeeman-splitting which for typical applied fields is
much smaller than the second term in Eq.(\ref{qp_disp}), the
so-called Doppler-shift, as well as any effects of the
supercurrent on $\Delta_{\bf k}$, $u_{\bf k}$ and $v_{\bf k}$
\cite{com1}.

Spin relaxation due to the ASF mechanism is described by the
spin-fermion model \cite{Morr98} where the spin susceptibility, $\chi$, is
renormalized by the interaction with low-energy fermionic degrees
of freedom and given by
\begin{equation}
\chi^{-1} = \chi_0^{-1} -  \Pi  \ .
\label{Dyson}
\end{equation}
Here
\begin{equation}
\chi_0^{-1}= { \xi_0^{-2} + ({\bf q} - {\bf Q}_i)^2 \over \alpha }
\ , \label{chi0}
\end{equation}
is the bare propagator, $\xi_0$ is the {\it bare}  magnetic
correlation length, $\alpha$ is a temperature independent
constant, ${\bf Q}=(\pi,\pi)$ is the position of the magnetic peak
in momentum space \cite{Morr00b} and $\Pi$ is the bosonic
self-energy given by the irreducible particle-hole bubble. A
supercurrent affects $\chi$ only through $\Pi$, which is
calculated using second order perturbation theory in the
spin-fermion coupling, $g$, and one obtains
\begin{equation}
\Pi=g^2 \Sigma \  ,
\end{equation}
with $\Sigma$ in Eq.(\ref{Sigma}). In the limit $\omega
\rightarrow 0$, Re$\, \Pi$ is only weakly momentum dependent and
can thus be included in a renormalized but momentum independent
correlation length, $\xi$, in Eq.(\ref{chi0}).

For both the ESF and ASF mechanism, we are thus left with the
calculation of Im$\, \Sigma$ which for $\omega \rightarrow 0$ is
dominated by particle-hole excitations in the vicinity of the
superconducting nodes. Expanding $E_{\bf k}$ up to linear order in
momentum around the nodes, we can perform the momentum and
frequency integrations in Eq.(\ref{Sigma}) analytically. Combining
the resulting expression for Im$\, \Sigma$ with Eqs.(\ref{T1T}),
(\ref{Dyson}) and (\ref{chi0}) we obtain the spin-lattice relaxation rate
presented in Eq.(\ref{T1T_full}) of the appendix. For our subsequent
discussion of the ESF and ASF mechanism it is necessary to briefly
review the form of $1/T_1$ in two limits. A low temperatures where
$|D_m/T| \gg 1$ for $m=1..4$, one obtains from Eq.(\ref{T1T_full})
\begin{eqnarray}
{1 \over T_1 T} &=& { A \over N} \left({ 1  \over v_F
v_\Delta }\right)^2 \sum_{n,m}  \Bigg\{ {\cal F}({\bf q}_{n,m})
\left(D_m \, D_n+{ \pi^2 T^2\over 3} \right) \nonumber \\ & &
\hspace{-0.5cm} + {\cal F}({\bf q}_{n,m+2}) \left(D_m \, D_n-{
\pi^2 T^2\over 3} \right) \Bigg\} +O(e^{D_m/T})
\label{T1T_lt}
\end{eqnarray}
where ${\cal F}({\bf q}_{n,m}),D_m$ are defined in the appendix,
and the sum runs only over those nodes with $D_m,D_n<0$. Since the
supercurrent in general breaks the lattice symmetry, the O(2) and
O(3) relaxation rates are expected to be different, as we shall
discuss below. In contrast, in the high-temperature limit
($|D_m/T| \ll 1$ for $m=1..4$) one has
\begin{equation}
{1 \over T_1 T} = { \pi A  \over 3N} \left( {T \over v_F v_\Delta
} \right)^2 \sum_{n,m=1}^4 {\cal F}({\bf q}_{n,m}) \label{T1T_ht}
\end{equation}
and we recover the temperature dependence of the relaxation rate
in the absence of a supercurrent.

A site specific relation between the resonance frequency
$\Delta\nu({\bf r})= \gamma_n \hbar H_{loc}({\bf r})$ where
$\gamma_n$ is the $^{17}$O nuclear gyromagnetic ratio and
$H_{loc}({\bf r})$ is the local magnetic field, and $1/T_1$ is
obtained via the local supercurrent momentum ${\bf p}_s({\bf r})
\sim \nabla \times {\bf H}_{loc}({\bf r})$; a relation which is
valid even in the presence of non-local as well as non-linear
corrections \cite{Amin98}. Following MW, we restrict our
discussion of the relaxation rates to nuclei which are further
than $R>2\xi_{ab}$ from the center of the vortex core (where
$\xi_{ab}$ is the superconducting in-plane coherence length)
\cite{com2}. We consider a hexagonal vortex lattice
\cite{Mag95,com6} and use the parameter set $v_F \approx 0.4$ eV,
$v_\Delta \approx 20$ meV, and $\xi \approx 2$ \cite{com5}, which
was shown to describe fermionic and magnetic excitations in
YBa$_2$Cu$_3$O$_7$ \cite{Morr00a,Morr00b}.
\begin{figure} [t]
\begin{center}
\leavevmode
\epsfxsize=6.5cm
\epsffile{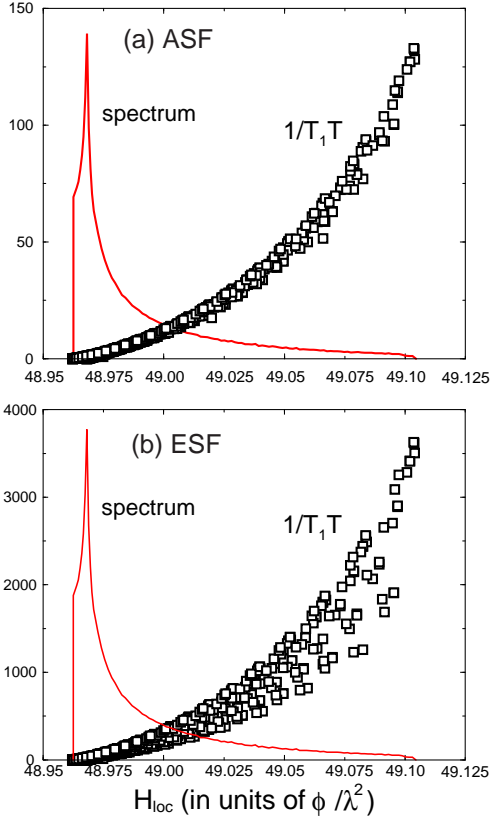}
\end{center}
\caption{$^{17}$O  spectrum (solid line) together with
$^{17}$O(2,3) $(1/T_1T)$ as a function of local magnetic field,
$H_{loc}$, for $T=1$K. {\it (a)} ASF mechanism, and {\it (b)} ESF
mechanism.} \label{17O_spec}
\end{figure}

In Fig.~\ref{17O_spec} we plot the $^{17}$O spectrum
\cite{Morr00a} which describes the distribution of local magnetic
fields (solid line) and $^{17}$O(2,3) $1/T_1T$ (open squares) for
{\it (a)} the ASF mechanism, and {\it (b)} the ESF mechanism as a
function of $H_{loc}$, i.e., resonance frequency. Nuclei at the
highest frequencies are located at a distance $2\xi_{ab}$ from the
center of the vortex (nuclei closer to the vortex core have been
omitted, see above discussion), nuclei at the lowest frequencies
are in the center of a vortex triangle, those at the maximum of
the spectrum are at the midpoint between two vortices.

For both relaxation mechanisms, $1/T_1T$ increases with increasing
resonance frequency, i.e., decreasing distance from the vortex
core, similar to the results obtained in
Refs.~\cite{Tak99,Morr00a,Wor00} for $^{63}$Cu. For $T=1$ K, the
relaxation rate for practically all nuclei is given by $1/T_1T
\sim p_s^2$, Eq.(\ref{T1T_lt}), and since $|{\bf p}_s|$ increases
with decreasing distance from the vortex core, $1/T_1T$ increases
consequently. However, $1/T_1T$ for the ESF mechanism exhibits a large
distribution of values for a given $H_{loc}$, in contrast to
$1/T_1T$ for ASF relaxation. Nuclei with the same resonance
frequency can in general possess different relaxation rates, since
$1/T_1T$ depends not only on the magnitude of ${\bf p}_s$, but also
on the angle between the direction of the local supercurrent and
the crystal axes, i.e., the Fermi velocity $v_F$ at the nodal
points, Eq.(\ref{T1T_lt}). It is exactly this angular dependence
which for ESF relaxation leads to the large distribution of values
for $1/T_1T$. The absence of such a large distribution
for the ASF mechanism is due to the antiferromagnetic enhancement
factor (i.e., the Stoner enhancement) in the denominator of
Eq.(\ref{calF_AF}).
\begin{figure} [t]
\begin{center}
\leavevmode
\epsfxsize=5.0cm
\epsffile{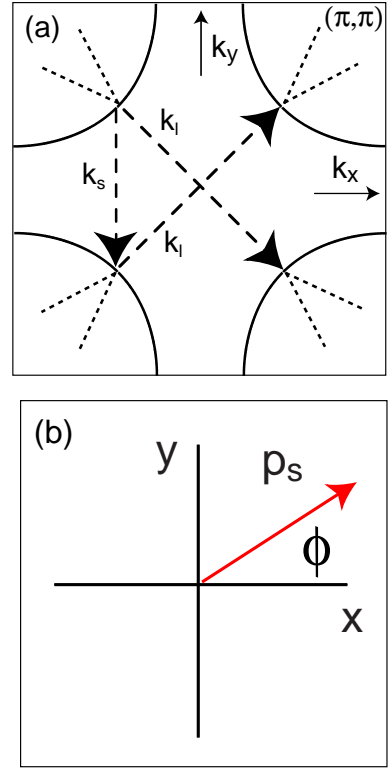}
\end{center}
\caption{{\it (a)} Fermi Surface and particle-hole excitations
(dashed arrow) with $q_{n,m}={\bf k}_{l,s}$ (nodes of the
superconducting gap are shown as dotted lines). {\it (b)} Angle
between ${\bf p}_s$ and the crystal x-axis.} \label{FS}
\end{figure}
The Stoner enhancement favors particle-hole excitations which
connect nodes at opposite sites of the Fermi surface (FS), i.e.,
with large momentum transfer, $q_{n,m}={\bf k}_{l}$ (see
Fig.~\ref{FS}a). Due to this enhancement, the second term on the
r.h.s.~of Eq.(\ref{T1T_lt}) yields the dominant contribution to
$1/T_1T$ for ASF relaxation in the low-temperature limit, and one
obtains
\begin{equation}
{1 \over T_1 T}  \approx  { A \over N} \left({ 1  \over v_F \,
v_\Delta }\right)^2 {\cal F}({\bf k}_{l}) \ \left( {\bf v}_F^2
{\bf p}_s^2 - 2 \pi^2 T^2/3 \right) \ .
\label{T1T_lt2}
\end{equation}
Thus the leading temperature contribution to $1/T_1T$ is
independent of the angle between ${\bf p}_s$ and the underlying
lattice, and the narrow distribution of $1/T_1T$ seen at higher
frequencies for ASF relaxation arises solely from subleading
contributions including particle-hole excitations with small
momentum transfer, e.g., with $q_{n,m}={\bf k}_{s}$ (see
Fig.~\ref{FS}a).

\begin{figure} [t]
\begin{center}
\leavevmode
\epsfxsize=7.5cm
\epsffile{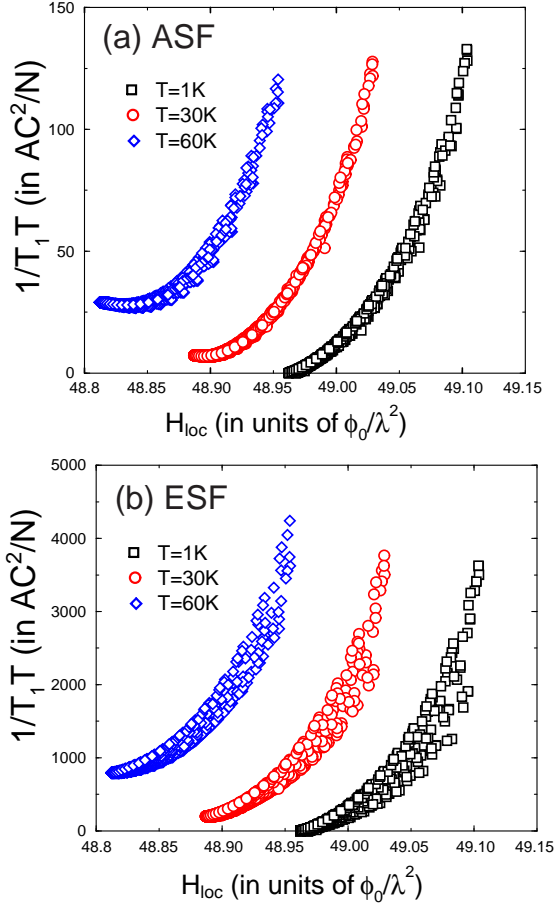}
\end{center}
\caption{$^{17}$O(2,3) $(1/T_1T)$ as a function of local field for
three different temperatures. {\it (a)} ASF mechanism, {\it (b)}
ESF mechanism. The 30 K and 60 K curves are horizontally offset
for clarity. }
\label{T1Tcomp}
\end{figure}

In Fig.~\ref{T1Tcomp} we present the $^{17}$O(2,3) relaxation rate
for three different temperatures (while the 30 K and 60 K curves
are horizontally offset for clarity, the left end points of all
curves correspond to the same local field at a distance
$2\xi_{ab}$ from the vortex core).
The temperature dependence of $1/T_1T$ in the ASF approach
(Fig.~\ref{T1Tcomp}a) is similar to that of $^{63}$Cu shown in
Fig.~2 of Ref.~\cite{Morr00a}. With increasing temperature,
$1/T_1T$ at high resonance frequencies decreases, in agreement
with Eq.(\ref{T1T_lt2}), while at low frequencies $1/T_1T \sim
T^2$, Eq.(\ref{T1T_ht}), and the relaxation rate increases. As a
result, $1/T_1T$ possesses a non-monotonic dependence on the local
field, as shown by the minimum in the relaxation rate for
intermediate frequencies, which was discussed in detail in
Ref.~\cite{Morr00a}.

In contrast, $1/T_1T$ for the ESF mechanism increases with
increasing temperature at all resonance frequencies. At low
frequencies, i.e., in the high temperature limit, this follows
directly from Eq.(\ref{T1T_ht}). At higher frequencies, i.e., in
the low temperature limit, an analysis of Eq.(\ref{T1T_lt}) shows
that due to the $^{17}$O form factors, the first term on the
r.h.s.~of Eq.(\ref{T1T_lt}) dominates  $1/T_1T$ which therefore
also increases with increasing temperature. Consequently, $1/T_1T$
does {\it not} exhibit a minimum as a function of the local field
as was the case for ASF relaxation. Note also that the
distribution of values for $1/T_1T$ decreases as the temperature increases,
since in the high temperature limit the relaxation rate becomes
independent of resonance frequency, i.e., $1/T_1T \sim T^2$. The
spin relaxation rate thus possesses {\it qualitatively} different
frequency and temperature dependencies for the ASF and ESF
mechanism, the origin of which lies in the antiferromagnetic
enhancement factor in the denominator of Eq.(\ref{calF_AF}) which
is only present for ASF relaxation.

\begin{figure} [t]
\begin{center}
\leavevmode
\epsfxsize=7.5cm
\epsffile{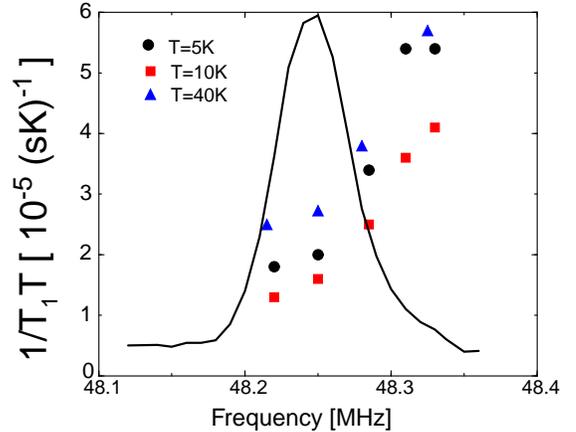}
\end{center}
\caption{Experimental data for the $^{17}$O spectrum (solid line) together
with $(1/T_1T)$ as a function of resonance frequency for three different
temperatures. Data are taken from Ref.\protect\cite{Cur00}.}
\label{ExpT1T}
\end{figure}

We next compare our theoretical results with the experimental data
on $^{17}$O(2,3) $1/T_1T$ by CMHS \cite{Cur00}, which we present
in Fig.~\ref{ExpT1T}.
The experimentally measured $1/T_1T$ increases with increasing
resonance frequency, in agreement with our theoretical results in
Fig.~\ref{T1Tcomp}, decreases between T=5K and T=10K with a larger
reduction at higher frequencies, and increases between T=10K and
T=40K. No evidence for a distribution of relaxation rates at high
frequencies was found so far, however, a further quantitative
analysis is still required to determine an upper bound for the
spread in $1/T_1$ \cite{Cur_pc}. Though both experimental features
agree qualitatively with our theoretical predictions for $1/T_1T$
in the ASF approach (see Fig.~\ref{T1Tcomp}a), spin-diffusion,
impurity effects and vortex vibrations \cite{Dem94} can
potentially contribute to the spin-lattice relaxation and thus
complicate the analysis of the experimental data within the ASF or
ESF approach. While it was argued \cite{Wor00} that spin-diffusion
is strongly suppressed by the inhomogeneity of the magnetic field
in a vortex lattice, CMHS concluded from the different temperature
dependence of the apical and planar oxygen relaxation rates, that
for $T \geq 25$ vortex vibrations are irrelevant for the
relaxation of $^{17}$O; thus the relaxation arises solely from the
electronic/magnetic mechanisms discussed above. For $T \leq 25$,
the apical and planar oxygen relaxation rates exhibit the same
linear temperature dependence, however, the planar $1/T_1$ is
still considerably larger than the apical $1/T_1$. Though this
result does not exclude the possibility that vortex vibrations
contribute to the relaxation process below 25 K, it suggests that
the relaxation mechanism is still dominated by electronic/magnetic
excitations. Clearly, further measurements of $1/T_1T$ are
required to study the various relaxation mechanisms \cite{com3}.
However, to the extent that the effects of vortex vibrations and
impurities are negligible at low temperatures, the experimental
data in Fig.~\ref{ExpT1T} support a predominant ASF mechanism of
the spin relaxation in the mixed state.

Since the supercurrent in general breaks the symmetry of the
underlying lattice, one expects different relaxation rates for
O(2) and O(3) for a given direction of ${\bf p}_s$. An experiment
which is able to reveal this difference in the relaxation rates is
a nuclear quadrupole resonance (NQR) measurement in which the
direction of a uniform supercurrent is varied.
\begin{figure} [t]
\begin{center}
\leavevmode
\epsfxsize=7.5cm
\epsffile{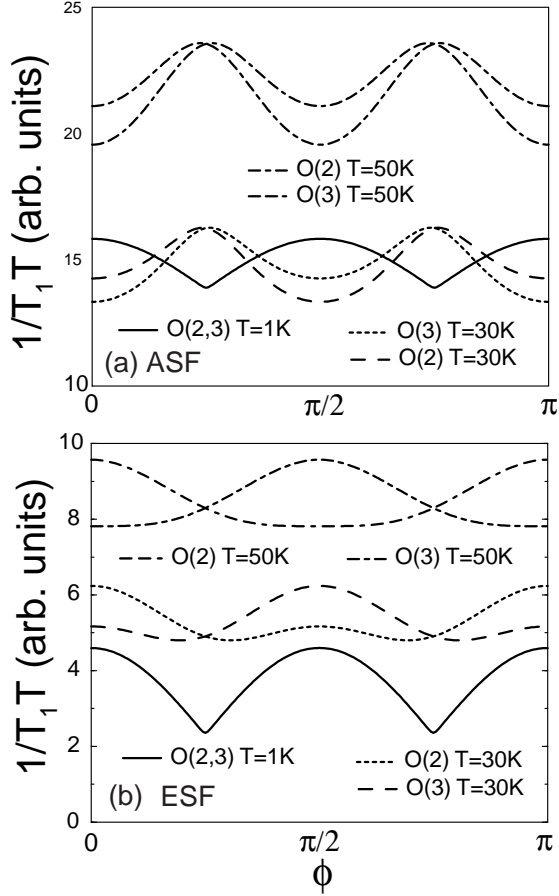}
\end{center}
\caption{ $^{17}$O(2,3)$1/T_1T$ for $p_s=0.02$ and three different
temperatures as a function of the angle $\phi$ between ${\bf p}_s$
and the crystal $x-$axis (see Fig.~\protect\ref{FS}b). {\it (a)}
ASF mechanism, {\it (b)} ESF mechanism. } \label{T1Tphi}
\end{figure}
In Fig.~\ref{T1Tphi}a we plot $1/T_1T$ for the ASF mechanism as a
function of the angle, $\phi$, between the supercurrent momentum,
${\bf p}_s$, and the crystal axes (see Fig.~\ref{FS}b). It follows
from Eq.(\ref{T1T_lt}) that in the limit $T \rightarrow 0$, O(2)
and O(3) $1/T_1T$ possess the same angular dependence. This
identical behavior again results from the Stoner enhancement,
since for the dominating particle-hole excitations with large
momentum transfer, ${\bf q}_{n,m}={\bf k}_l$, (see Fig.~\ref{FS}),
the form factors $F(q_{n,m})$ for O(2) and O(3) are the same. With
increasing temperature, the relaxation rates for O(2) and O(3)
begin to deviate, since excitations with smaller momentum transfer
become increasingly more important, however, the maxima and minima
of O(2) and O(3) $1/T_1T$ still occur at the same angle $\phi$.

The relaxation rates of O(2,3) for the ESF mechanism,
(Fig.~\ref{T1Tphi}b), are also identical in the low-temperature
limit, as follows directly from Eq.(\ref{T1T_lt}). However, with
increasing temperature, the angular dependence of $1/T_1T$ for
O(2) and O(3) is {\it out of phase}: the minima of O(2) $1/T_1T$
coincide with the maxima of O(3) $1/T_1T$. To demonstrate that
this ''phase-shift" originates from the different form factors of
O(2,3), Eq.(\ref{form}), we consider the case $\phi=\pi$. One then has from
Eq.(\ref{T1T_lt}) for O(2)
\begin{eqnarray}
{1 \over T_1 T}  &\sim&  [1+cos^2(k_l^x/2)]
\Big(D_1^2+D_4^2+2D_1D_4\big) \nonumber \\
& & +4 [1-cos^2(k_l^x/2)]\pi^2 T^2/3 \ ,
\end{eqnarray}
whereas for O(3) one obtains
\begin{eqnarray}
{1 \over T_1 T}  &\sim&  [1+cos^2(k_l^x/2)]
\Big(D_1^2+D_4^2+2D_1D_4\big) + O(T^4) \ .
\end{eqnarray}
For $T=0$ we indeed find that the O(2,3) relaxation rates are
identical, while for $T>0$ (and $\phi=\pi$) O(2) $1/T_1T$ is
larger than O(3)$1/T_1T$, in agreement with our numerical results
in Fig.~\ref{T1Tphi}.

Finally, while the overall scale of $1/T_1T$ within the ESF and
ASF approach depends on a variety of parameters, e.g., the
hyperfine coupling, $C$, the magnetic correlations length, $\xi$,
the detailed form of the Fermi surface, etc., we find that its
qualitative features are robust against changes in these
parameters. Thus our predictions for the qualitative differences
between the ESF and ASF relaxation mechanisms are valid for all
cuprate materials, and {\it do not} depend on any fine-tuning of
parameters.

In summary, we present a theoretical scenario for the spin-lattice
relaxation rate of $^{17}$O in the mixed state of the high-$T_c$
cuprates. We consider $1/T_1$ for two relaxation mechanisms: one
based on electronic spin-flip scattering of BCS-type electrons, and
one due to antiferromagnetic spin fluctuations. We show that the
predicted temperature and frequency dependence of $1/T_1$ differs
qualitatively for these two mechanisms. A comparison of our
theoretical results with the available experimental data from
Ref.~\cite{Cur00} suggests that the ASF mechanism dominates the
relaxation rate. To the extent that this conclusion is confirmed
by future more detailed experiments, it implies that the
relaxation rate of $^{63}$Cu in the mixed state is also dominated
by the ASF mechanism, in contrast to the scenarios pursued in
Refs.~\cite{Tak99,Wor00}. Finally, we propose an NQR experiment in
which the direction of a uniform supercurrent with respect to the
crystal lattice is varied and predict a unique angular dependence
of $1/T_1$ for O(2) and O(3).

It is our pleasure to thank N. Curro, W. Halperin, C. Milling, D.
Pines, J. Sauls, R. Wortis, and especially C. P. Slichter for valuable
discussions. This work has been supported by the Department of
Energy at Los Alamos National Laboratory.

\appendix
\section{}
\label{appendix}

The general form for $1/T_1T$ in the presence of a supercurrent is
given by
\begin{eqnarray}
{1 \over T_1 T} &=& { A \over N} \left({ 1  \over v_F
v_\Delta }\right)^2 \sum_{n,m}  \Bigg\{ {\cal F}({\bf q}_{n,m})
 \, G_1(D_m,D_n,T) \nonumber \\ & &
\hspace{-0.5cm} + {\cal F}({\bf q}_{n,m+2}) \, G_2(D_m,D_n,T) \Bigg\}
\label{T1T_full}
\end{eqnarray}
where $D_m={\bf v}^{(m)}_F \cdot {\bf p}_s$, ${\bf v}^{(m)}_F$ is
the Fermi velocity at the node in the $m$'th quadrant of the
Brillouin zone, $v_\Delta=|\partial \Delta_{\bf k} /\partial {\bf
k}|$ at the nodes, and ${\bf q}_{n,m}$ is the momentum connecting
nodes $n$ and $m$. For ESF scattering we find
\begin{eqnarray}
A &=&  k_B  (\hbar^2 \gamma_n \gamma_e)^2/(2 \hbar) \ ; \nonumber
\\
{\cal F}({\bf q}_{n,m}) &=& F_{O(2,3)}({\bf q}_{n,m}) \ ,
\label{calF_ES}
\end{eqnarray}
while for the ASF mechanism on has
\begin{eqnarray}
A &=& (\alpha g)^2 k_B  (\hbar^2 \gamma_n \gamma_e)^2/(2 \hbar) \
; \nonumber \\
{\cal F}({\bf q}_{n,m}) &= & {F_{O(2,3)}({\bf
q}_{n,m}) \over (\xi^{-2} + ({\bf q}_{n,m}-{\bf Q})^2)^2} \ .
\label{calF_AF}
\end{eqnarray}
In both cases,  $G_{1,2}$ is given by
\begin{eqnarray}
G_1(D_m,D_n,T)&=& {D_n \, D_m + \epsilon^2 \over T^2} \,  n_F(\epsilon)
 + {\pi^2\over 6}
\nonumber \\
& & \hspace{-1.5cm}+ \, {D_n+D_m \over T}  \Bigg( {\epsilon[1-n_F(\epsilon)]
\over T} +\ln[n_F(\epsilon)] \Bigg) \nonumber \\
& & \hspace{-1.0cm} -2 \int_0^{\epsilon/T} dx \, x \, n_F(x)
\label{G1}
\end{eqnarray}
where $n_F(\epsilon)=[\exp(\epsilon/T)+1]^{-1}$ is the Fermi function,
$\epsilon=max\{D_n,D_m\}$, and
\begin{eqnarray}
G_2(D_m,D_n,T)&=& {D_n \, D_m \over T^2} \left[n_F(D_m)-n_F(-D_n)
\right] \nonumber \\ & & \hspace{-2.0cm}+ \, {D_n-D_m \over T}
\Bigg\{ {D_n \, [n_F(-D_n)-1] \over T}+ \ln[n_F(-D_n)] \nonumber
\\ & &\hspace{-1.5cm} + {D_m \, [n_F(D_m)-1] \over T}-
\ln[n_F(D_m)] \Bigg\} \nonumber \\ & & \hspace{-2.0cm} +
\left({D_n \over T}\right)^2 \, n_F(-D_n)- \left({D_m \over
T}\right)^2 \, n_F(D_m) \nonumber \\ & & \hspace{-1.5cm} -2
\int_{D_m/T}^{-D_n/T} dx \, x \, n_F(x) \qquad {\rm for}\, \,  D_n+D_m \leq 0
\nonumber \\[0.2cm]
G_2(D_m,D_n,T)&\equiv&0 \qquad {\rm for}\, \, D_n+D_m>0 \ .
\label{G2}
\end{eqnarray}


\begin{thebibliography}{99}
\bibitem{com4} We show below that the qualitative difference between
ESF and ASF relaxation arises from the Stoner enhancement of the
spin susceptibility which is present for the ASF mechanism, but
absent for the ESF mechanism. Thus the temperature and frequency
dependence of $1/T_1$ obtained in Refs.\protect\cite{Tak99} and
\protect\cite{Wor00} agree qualitatively.
\bibitem{com6} Note that the authors of Ref.\protect\cite{Tak99} assumed a
square vortex lattice which is inconsistent with the experimental
results of Ref.\protect\cite{Mag95}.
\bibitem{com5} Since $\xi$ is temperature independent in the superconducting
state it does not contribute to the temperature dependence of
$1/T_1T$.
\end{thebibliography}
\end{document}